\documentclass[prb,twocolumn]{revtex4-1}

\usepackage[varg]{txfonts}

\usepackage{amssymb}

\usepackage{graphicx}
\usepackage{graphics}
\usepackage{bbold,bm}
\usepackage{hyperref}
\usepackage{epsfig}
\usepackage{dcolumn}

\newcommand\vex[1]{\mathbf{#1}}

\def\re{\mathrm{Re}\,}
\def\im{\mathrm{Im}\,}
\def\tr{\mathrm{Tr}}
\def\dd{\mathrm{d}}
\def\id{\mathbb{1}}
 
\def\dbar{\hbox{$d$\kern-0.6em\raise0.3em\hbox{$-$}}\hspace{-0.5mm}}

\begin{document}

\title{Topological exciton condensate of imbalanced electrons and holes}

\author{Babak Seradjeh}

\address{Department of Physics, Indiana University, 727 East Third Street, Bloomington, IN 47405-7105 USA}

\begin{abstract}
I study the effects of particle-hole imbalance on the exciton superfluid formed in a topological insulator thin-film and obtain the mean-field phase diagram. At finite imbalance a spatially modulated condensate is formed, akin to the Fulde-Ferrell-Larkin-Ovchinnikov state in a superconductor, which preempts a first-order transition from the uniform condensate to the normal state at low temperatures. The imbalance can be tuned by changing the chemical potential at the two surfaces separately or, alternatively, by an asymmetric application of Zeeman fields at constant chemical potential. A vortex in the condensate carries a precisely fractional charge half of that of an electron. Possible experimental signatures for realistic parameters are discussed.
\end{abstract}

\maketitle

%---- Introduction
\section{Introduction}
The Cooper pairs in superconductors known so far have zero center-of-mass momentum and, consequently, the order parameter is spatially uniform. In a singlet superconductor the two electrons in a pair carry opposite spin and superconductivity may be understood as an instability of their nested Fermi surfaces. The nesting is lost when the Fermi surfaces are different in size (e.g., due to the Zeeman interaction in a magnetic field). But the instability may still be recovered by translating the Fermi surfaces in momentum space so that they are partially nested. The pairs will then carry a finite center-of-mass momentum equal to the momentum shift and the order parameter will be spatially modulated, as first suggested theoretically by Fulde and Ferrell~\cite{FulFer64a} and, separately, by Larkin and Ovchinnikov~\cite{LarOvc64a} (FFLO). 

No conclusive experimental evidence for the FFLO state has been found so far.~\cite{CasNar04a} Recently, there has been an effort~\cite{LiaRitPap10a} in realizing the FFLO state in imbalanced two-species Fermi gases with magnetically tuned interaction between the two species. A different approach is via electron-hole bilayers where a neutral exciton superfluid is formed by the Coulomb interaction between the two layers. A double-layer graphene structure where electrons and holes are hosted on opposite surfaces in an external electric field is a candidate with a potentially high critical temperature.~\cite{MinBosSu08a,SerWebFra08a} Also, a thin film of a strong topological insulator (STI) was argued~\cite{SerMooFra09a} to realize a novel form of the exciton superfluid, dubbed topological exciton condensate (TEC), where the special topology of the bulk results in fractionally charged vortices and protection against weak disorder.~\cite{HasKan10a}
Recent material improvements~\cite{JiaSunChe12a} bode well for the experimental realization of TEC.  On the theoretical side, the realization of monopoles by vortices,~\cite{RosFra10a} their effective theory,~\cite{ChoMoo11a} and the effects of screening in orbitally coupled magnetic field~\cite{TilLeeHan11a} have been studied. But these studies only consider a uniform condensate, requiring the mean surface chemical potential to be fine-tuned to the Dirac node (hereafter set as zero of energy) to obtain electron-hole balance.

%---Fig 1
\begin{figure}
%\begin{center}
\centerline{\includegraphics[scale=1]{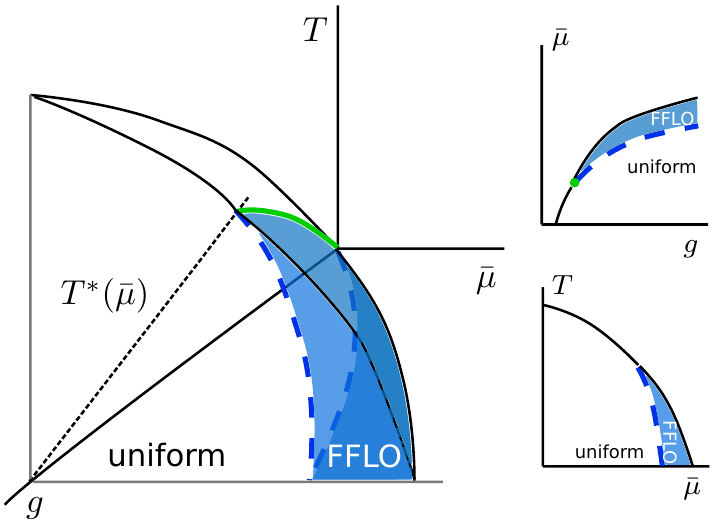}}
%\end{center}
\caption{(color online) The mean-field phase diagram. The axes are the interaction strength $g$, temperature $T$, and the mean chemical potential of the two surfaces $\bar\mu$. The solid (dashed) lines indicate a continuous (first-order) phase transition. The nonuniform state is labeled FFLO and shaded (blue). The dotted line separates the continuous and first-order transitions at a critical line (thick green).  The insets show projections on constant $T$ (top) and $g$ (bottom) planes.}\label{fig:pd}
\end{figure}

In this work, I study the effect of electron-hole imbalance on the TEC. I show that a FFLO state is realized for a finite electron-hole imbalance. The imbalance can be achieved by tuning the mean surface chemical potential away from zero by external gates or, as I show in this paper, also by an asymmetric Zeeman field normal to the layers. Unlike the uniform state, the gap equation for the FFLO state in a STI thin-film is different from the superconducting analog or the semiconductor bilayers due to the spinor structure of the Dirac dispersion of the STI surface states. However, for small imbalance, a similar phenomenology is obtained: there is a first-order transition from the uniform to the FFLO condensate with the wave vector of spatial modulations set by the particle-hole imbalance at low temperatures, followed by a continuous transition to the normal state. The results are summarized in the phase diagram shown in Fig.~\ref{fig:pd}. I also show that a vortex in the FFLO condensate carries a fractional charge. 

This study suggests that the FFLO condensate is generically present when the electrons and holes on the two surfaces are slightly imbalanced. Therefore, the topological insulator thin film is a playground where the FFLO state inherits the topological features of the parent structure, making it unique among current proposals. It also offers several control knobs (separate surface chemical potentials, film thickness, and Zeeman fields) to explore excitonic condensates and fractionalization in the absence of complications arising from the orbital effects of the magnetic field.

%---- Gap equation
\section{Generalized gap equation}
In order to describe an inhomogeneous condensate, I define a four-spinor 
%%%%
$$
{\Psi_{\vex p}^{T}(\vex k)} = \left(\psi_1^{T}(\vex k+\frac12\vex p), \psi_2^{T}(\vex k -\frac12 \vex p)\right),
$$
%%%%
where $\psi_{\alpha}=(\psi_{\alpha\uparrow},\psi_{\alpha\downarrow})$ is the two-spinor electron annihilation operator at surface $\alpha=1,2$ and spin projection $\uparrow,\downarrow$ normal to the surface, and $\vex p$ is the wave vector of the spatial modulations of the condensate order parameter. Then, the noninteracting Hamiltonian of the two surfaces of the STI is $H_0 = \sum_{\vex k} \Psi_{\vex p}(\vex k)^\dagger h_0 \Psi_{\vex p}(\vex k)$, with
%%%%
\begin{equation}\label{eq:h0}
h_0 = v_F \tau_z \boldsymbol{\sigma}\cdot\vex k + \frac12 v_F \boldsymbol{\sigma}\cdot\vex p-\Delta\mu\tau_z-\bar{\mu},
\end{equation}
%%%%
where $\boldsymbol{\tau}$ and $\boldsymbol{\sigma}$ are Pauli matrices acting at the surface and spin space, respectively, $\Delta\mu=\frac12(\mu_1-\mu_2)$ and $\bar{\mu}=\frac12(\mu_1+\mu_2)$ with the chemical potential $\mu_\alpha$ at surface $\alpha$. In a continuum description the sums are replaced by integrals appropriately. I set $\hbar=k_B=1$ throughout.

The Coulomb interaction has both an intralayer and an interlayer component. I will not consider the intralayer interaction explicitly, assuming its effects are taken into account by renormalizing the Fermi velocity. The interlayer Coulomb interaction is  
%%%%
\begin{equation}
U = \frac1N \sum_{\vex k} g_{\vex k} n_1(\vex k)n_2(-\vex k),
\end{equation}
%%%%
where $n_\alpha=\psi_\alpha^\dagger\psi_\alpha$ is the electron density in surface $\alpha$, $N$ is the number of sites, and $g_{\vex k}=2\pi e^2 \exp(-k d)/\varepsilon\ell^2 k$, with $\ell$ the lattice spacing, $d$ the distance between the layers, and $\varepsilon$ the dielectric constant of the intermediate medium.

I will now derive the mean-field Hamiltonian describing the exciton condensate. The order parameter is
%%%%
\begin{equation}
m_{\vex q,\vex p} = \frac1N \sum_{\vex k} g_{\vex k-\vex q} \langle \psi_1(\vex k+\vex p) \psi_2^\dagger(\vex k) \rangle.
\end{equation}
%%%%
The mean-field Hamiltonian is then $H_{\mathrm{MF}} = E_0 + \sum_{\vex k} \Psi_{\vex p}^\dagger(\vex k) h(\vex k) \Psi_{\vex p}(\vex k)$ where the reduced Hamiltonian $h$ is
%%%%
\begin{equation}\label{eq:h}
h = h_0 + \re(m_{\vex k,\vex p}) \tau_x - \im(m_{\vex k, \vex p}) \tau_y,
\end{equation}
%%%%
and
$
E_0 = \sum_{\vex r} \tr\left[ m_{\vex p}(\vex r) g^{-1}(\vex r) m_{\vex p}^\dagger(\vex r) \right]
$
(the trace is over spin) is the kinetic energy associated with the order parameter $m_\vex p(\vex r)\propto g(\vex r) \sum_{\vex x} \langle \psi_1(\vex x) \psi_2^\dagger(\vex x-\vex r) \rangle e^{-i\vex p\cdot\vex x}$. 

%---Fig 2
\begin{figure}[t]
%\begin{center}
\centerline{\includegraphics[width=3.3in]{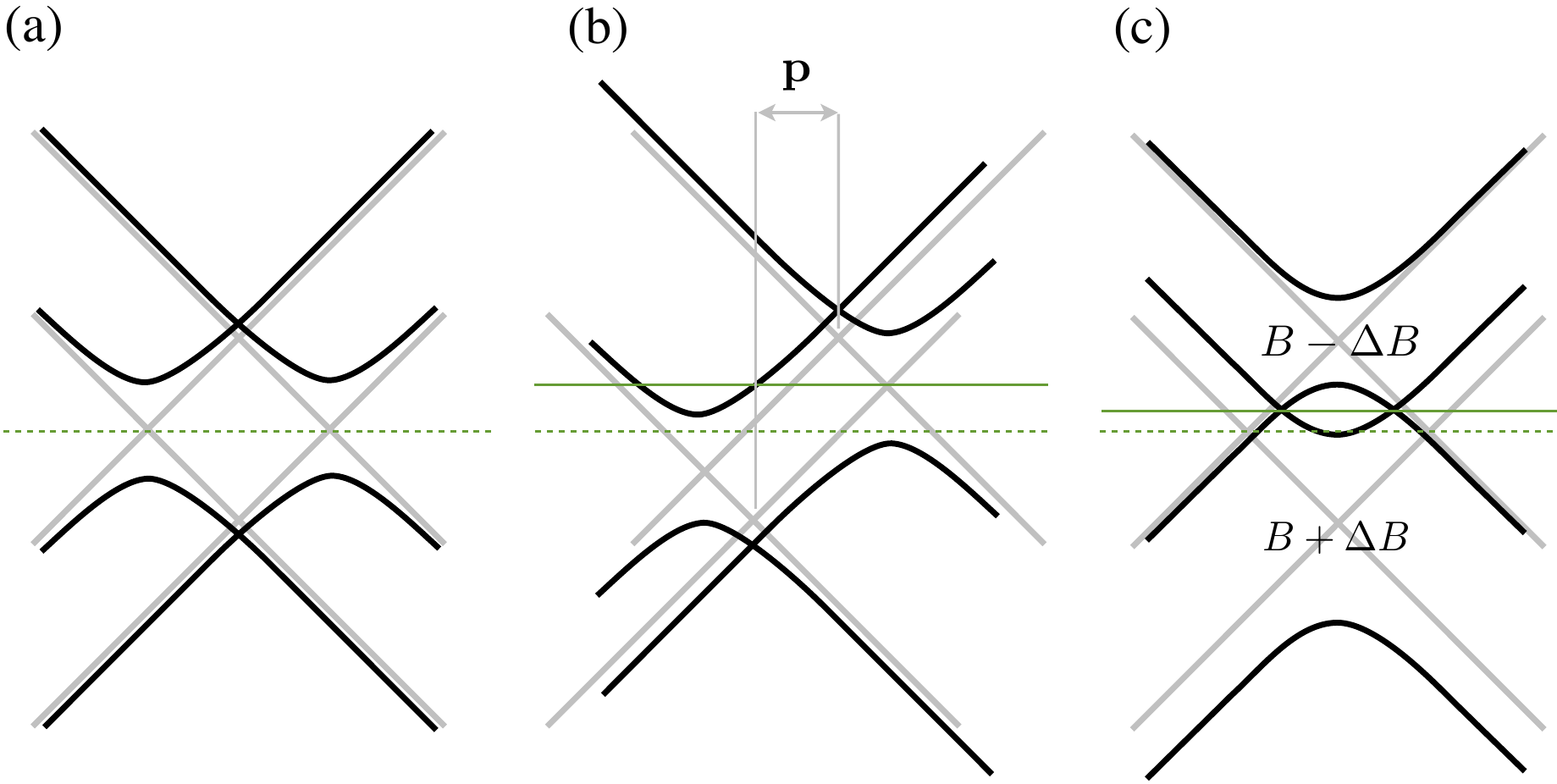}}
%\end{center}
\caption{(color online) The spectrum of energy (vertical) vs. momentum (horizontal). The black (gray) solid lines show the spectrum for finite (zero) $m$. The dashed (green) horizontal line shows the chemical potential at the Dirac nodes for reference. (a) $\vex p=0$; the particle and hole Fermi surfaces are fully nested.  (b) $\vex p\neq0$. The two Dirac cones in each surface are shifted by $\pm\frac12\vex p$; the solid (green) horizontal line marks the chemical potential $\bar\mu=\frac12v_F|\vex p|$ for which partial nesting between particle and hole Fermi surfaces is obtained.  (c) $B, \Delta B\neq0$ and $\vex p=0$; the solid (green) horizontal line marks the chemical potential $\bar\mu=B\Delta B/\Delta\mu$ where particle-hole symmetry is obtained.}\label{fig:spec}
\end{figure}

The gap equation is found by minimizing the free energy $F=-T\log\langle e^{-H_{\mathrm{MF}}/T} \rangle$ with respect to $m_{\vex q, \vex p}$. Assuming $m_{\vex k,\vex p}\propto\id$ in spin space~\cite{SerMooFra09a}, the result at $T=0$ is
%%%%
\begin{equation}
m_{\vex q,\vex p} = -\frac1{4N} {\sum_{\vex k s}}' \frac{\left(E_{\vex k s,\vex p}^2-\epsilon_{\vex k,\vex p}^2 + \frac12 v_F^2 \vex p^2\right)g_{\vex k-\vex q} m_{\vex k,\vex p}}{E_{\vex ks,\vex p}(E_{\vex ks,\vex p}^2-\epsilon_{\vex k,\vex p}^2)+v_F^2\Delta\mu\,\vex p\cdot \vex k}.
\end{equation}
%%%%
Here $E_{\vex k s,\vex p}$ is the eigenenergy of the reduced Hamiltonian~(\ref{eq:h}) with momentum $\vex k$ and spin/surface index $s$, $\epsilon_{\vex k,\vex p}^2 = v_F^2(\vex k^2 + \frac14 \vex p^2) + \Delta\mu^2 + |m_{\vex k,\vex p}|^2$, and the sum is restricted to occupied states with $E_{\vex ks,\vex p}<\bar\mu$. The analytical form of $E_{\vex ks,\vex p\neq0}$ is complicated, but it is straightforward to visualize, as shown in Figs.~\ref{fig:spec}(a) and~\ref{fig:spec}(b).

In the following, I will assume for simplicity a constant $g_{\vex k}=g$. Then $m_{\vex q,\vex p}=m_{\vex p}$ independent of $\vex q$. Passing to finite temperature by introducing the Fermi-Dirac distribution function $n_{F}(z) = (1+e^{z/T})^{-1}$,  the gap equation reads
%%%%
\begin{equation}\label{eq:gapeq-g}
1=-\frac{g}{4N} {\sum_{\vex k s}^\Lambda} \frac{\left(E_{\vex ks,\vex p}^2-\epsilon_{\vex ks,\vex p}^2 + \frac12 v_F^2 \vex p^2\right)n_{F}(E_{\vex ks,\vex p}-\bar\mu)}{E_{\vex ks,\vex p}(E_{\vex ks,\vex p}^2-\epsilon_{\vex ks,\vex p}^2)+v_F^2\Delta\mu\,\vex p\cdot \vex k},
\end{equation}
%%%%
where $\Lambda$ is a momentum cutoff. This is one of the main results of this paper.

First, let us note that for $\vex p =0$ we recover the gap equation in Ref.~\onlinecite{SerMooFra09a}. Second, I note that the functional form of the gap equation is different when $\vex p\neq0$. This is caused by the distinct spinor structure of the surface Hamiltonian and is in contrast to the superconducting case where the FFLO state is governed by the same gap equation as the uniform state. However, for small $v_F|\vex p|/m_0$ (where $m_0$ is the uniform condensate at $T=0$) the deviations are small and we obtain a similar phenomenology.

The change in the free energy by the condensate may be written as~\cite{Abr88a}
%%%%
\begin{equation}\label{eq:DF}
\Delta F \equiv F(m) - F(0) =  - \int_0^{m} \frac{\widetilde m^2}{g^2} \frac{\partial g}{\partial \widetilde m} \dd \widetilde m, % = \int_0^{g}d g' m^2/g'^2 =
\end{equation}
%%%%
where $g=g(m,\bar\mu,T)$ is calculated from the gap equation. Therefore, a positive (negative) slope $\partial g/\partial m$ lowers (increases) the free energy.

%---Fig 3
\begin{figure}[tb]
%\begin{center}
\centerline{\includegraphics[scale=1]{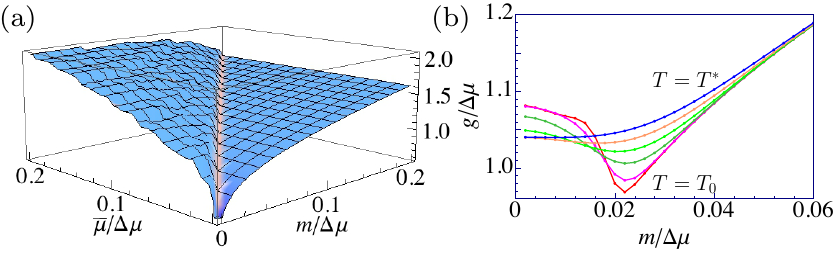}}
%\end{center}
\caption{(color online) Uniform solution, $\vex p=0$ ($v_{F}\Lambda/\Delta\mu=2$ and units are $\ell\Delta\mu/v_F=2$). (a) The solution $g(m,\bar\mu,T=0)$. Clearly, $g_{\mathrm{min}}$ is at $m=\bar\mu$. (b) The dependence of $g$ on temperature $\{ T_0/\Delta\mu=0.001, 0.002, 0.004, 0.006, 0.008, T^*/\Delta\mu=0.01\}$ for $\bar\mu/\Delta\mu=0.02$.}\label{fig:uniform}
\end{figure}

%---- Uniform condensate: 1st order phase transition
\section{Uniform condensate}
The dependence of $g(m,\bar\mu,T)$ on different parameters for $\vex p=0$ is shown in Fig.~\ref{fig:uniform}. At $T=0$ the minimum, $g_{\mathrm{min}}$, clearly happens at $m=\bar\mu$. At constant $\bar\mu$, the free energy $\Delta F$ in Eq.~(\ref{eq:DF}) is lowered (increased) by $g(m)$ for $m>\bar\mu$ ($m<\bar\mu$). As a result, for a given $g$, there is a finite value $\bar\mu=\bar\mu_{c1}$, $m=m_{c1}>\bar\mu_{c1}$ where these contributions cancel each other and there is a first-order transition to a normal state.

At $T>0$ the dip in $g(m,\bar\mu,T)$ is rounded and $g_{\mathrm{min}}$ increases with $T$. The upturn for $m<\bar\mu$ is also rounded and the value $g(0,\bar\mu,T)$ initially decreases with increasing $T$. As a result, $g_{\mathrm{min}}(\bar\mu,T)\to g(0,\bar\mu,T)$ continuously as $T$ approaches a value $T^*(\bar\mu)$. For $T>T^*(\bar\mu)$, $g$ increases monotonically with $m$. For a given $g=g_0$, let us denote by $T_c(\bar\mu)$ the temperature at which $g(0,\bar\mu,T)$ goes from below to above $g_0$. If $T^*(\bar\mu)<T_c(\bar\mu)$, the transition to the normal state will be continuous. However, if $T^*(\bar\mu)>T_c(\bar\mu)$, the transition will be first-order, since before reaching $T_c$ the gain in free energy will be depleted by the upturn in $g(m)$ at some temperature $T_{c1}(\bar\mu)<T^*(\bar\mu)$. We can also see this as follows. Take  $g^*(\bar\mu)\equiv g\left[0,\bar\mu, T^*(\bar\mu)\right]$. Given $g=g_0$, there is $\bar\mu^*(g_0)$ where $g^*(\bar\mu^*)=g_{\mathrm{min}}\left(\bar\mu^*,T^*(\bar\mu^*)\right)=g_0$. For $\bar\mu<\bar\mu^*$, we have $g^*(\bar\mu)<g_0$ and therefore the transition will be continuous at some $T_c(\bar\mu)>T^*(\bar\mu)$. For $\bar\mu>\bar\mu^*$, however, $g^*(\bar\mu)>g_0$ and the transition will turn first order at $T_{c1}(\bar\mu)<T^*(\bar\mu)$.

%---Fig 4
\begin{figure}[tb]
%\begin{center}
\centerline{\includegraphics[scale=1]{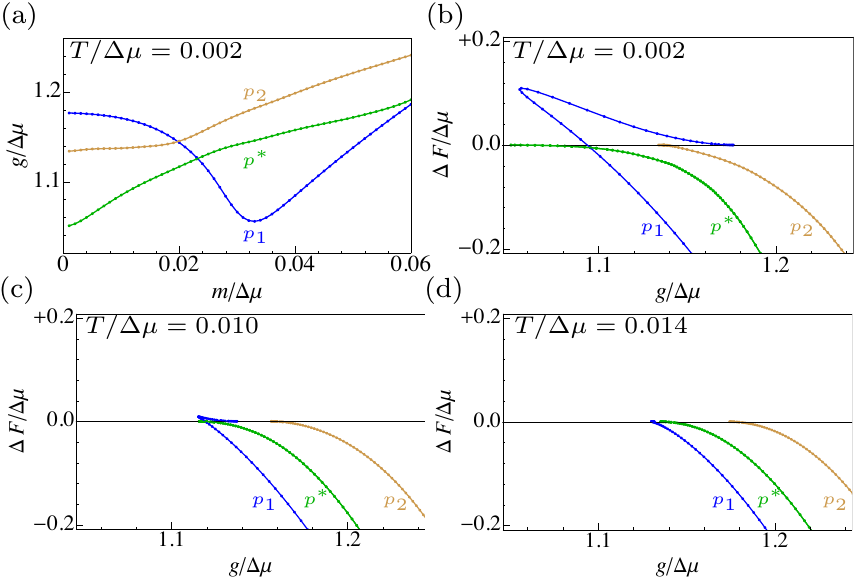}}
%\end{center}
\caption{(color online) FFLO solution for $\bar\mu/\Delta\mu=0.03$ ($\Lambda$ and units are as in Fig.~\ref{fig:uniform}).  (a) $g(m,\bar\mu,T)$ for $\vex p=(p_x,0)$, $v_Fp_x/\Delta\mu=\{0.002, 0.06, 0.1\}$ where $p^*=2\bar\mu/v_F$ gives the largest partial nesting between the electron and hole Fermi surfaces in Fig.~\ref{fig:spec}(b).  (b), (c), (d) The change in free energy $\Delta F$ for different temperatures. The first-order transition from the uniform to the FFLO state for $T<T^*$ in (b) and (c) becomes second-order for $T>T^*$ in (d) and directly to the normal state.}\label{fig:FFLO}
\end{figure}

%---- FFLO
\section{FFLO condensate}
The typical $g(m_{\vex p},\bar\mu,T\approx0)$ for fixed $\bar\mu$ and various values of $\vex p$ is plotted in Fig.~\ref{fig:FFLO}(a). The free energy  change $\Delta F$ is shown in Figs.~\ref{fig:FFLO}(b), \ref{fig:FFLO}(c), \ref{fig:FFLO}(d) as a function of $g$, showing that the first-order transition from the uniform condensate to the normal state found above is pre-empted by a first-order transition to the FFLO state with $\frac12 v_F|\vex p^*|=\bar\mu$ at $T=0$, followed by a continuous transition from the FFLO state to the normal state. For $T>T^*$, there is only a continuous transition from the uniform condensate to the normal state. At finite $T$, the wave vector of the spatial modulations of the FFLO condensate acquires temperature dependence and eventually vanishes at the critical point $T=T^*$.

Overall, we obtain the phase diagram shown in Fig.~\ref{fig:pd}, where the FFLO state is shaded. I note that this analysis is performed assuming an isotropic dispersion. As a result the states with the same $|\vex p|$ are degenerate. The ground state is then a superposition of such states, chosen by the residual interactions in the system or the anisotropy of the dispersion at higher momenta.

It was previously shown~\cite{SerMooFra09a} that a vortex in the uniform condensate binds a fractional charge $\frac12e$ due, at the mean-field level, to a zero-energy bound state, $(h+\bar\mu)\Psi_0=0$, at the vortex core~\cite{SerWebFra08a}. The spectrum is symmetric for all $\vex p$ since $\Gamma (h+\bar\mu) \Gamma^\dagger = -(h+\bar\mu)$, where $\Gamma=\tau_y\sigma_y K$ and $K$ is the complex conjugation. (Note that in real space $\vex k\to-i\boldsymbol{\nabla}$ in Eq.~(\ref{eq:h0}) but $\vex p$ is a fixed parameter.) That is, all states at nonzero energies must come in pairs. So, the single zero-energy state found for $\vex p=0$ persists when $\vex p\neq0$ and a vortex in the FFLO condensate also binds a fractional charge $\frac12e.$

%---- Magic field
\section{Zeeman field}
A Zeeman field normal to surface $\alpha$ enters the Hamiltonian~(\ref{eq:h}) via a term $B_\alpha \sigma_z$, where $B_\alpha$ is the Zeeman energy at the surface $\alpha$. This can be absorbed into Eq.~(\ref{eq:h0}) by replacing $\vex k$ and $\vex p$ with the three-vectors $k=(\vex k, \Delta B/v_F )$ and $p=(\vex p, 2B/v_F)$ where $\Delta B = \frac1{2}(B_1-B_2)$ and $B =\frac12(B_1+B_2)$. With this replacement, the gap equation in Eq.~(\ref{eq:gapeq-g}) remains the same.

Due to the linear density of states of the Dirac cone, electron-hole balance is restored for
%%%%
\begin{equation}
B\Delta B=\bar\mu\Delta\mu.
\end{equation}
%%%%
See Fig.~\ref{fig:spec}(c) for a visual representation. Therefore, in principle, the transition to the FFLO condensates can also be fine-tuned by an asymmetric Zeeman field between the surfaces. For instance with $B=\Delta B = 10^{-2}$~meV, and $\Delta\mu\sim1$--$10$~meV one can achieve $\bar\mu/\Delta\mu\sim10^{-4}$--$10^{-6}$.

%---- Discussion and conclusions
\section{Discussion}
Let us make some estimates of the parameters. In the prototypical topological insulator Bi$_2$Se$_3$ the bulk insulating behavior is obtained when the thin film is a few quintuple layers, say, $d\sim10$~nm. Assuming $\varepsilon\sim30$ we find $g\sim5$~meV. The cutoff $\Lambda\sim1/d$, so taking $v_F\sim1$~eV\AA, we have $v_F\Lambda\sim10$~meV. Using $\Delta\mu\sim5$~meV we find~\cite{SerMooFra09a} $m_0=\sqrt{v_F\Lambda \Delta\mu}e^{-v_F^2\Lambda^2/g\Delta\mu}\sim0.1$~meV. We should expect the FFLO state to appear when $\bar\mu$ is near its value for the metastable first-order transition. We can estimate this from the critical interaction strength~\cite{SerMooFra09a} as $\bar\mu\lesssim \Delta\mu e^{-v_F\Lambda/\Delta\mu}\sim0.1$~meV, corresponding to a wavelength of modulations $\lambda_{\mathrm{FFLO}}\gtrsim500$~nm in the visible range.

I will now turn to possible experimental signatures. The exciton condensate can sustain counter propagating superflow in the two surfaces, manifested in a large Coulomb drag~\cite{VigMac96a}, whereby electric field in one surface induces a counterflow current in the other. The spatial profile of the current can distinguish the uniform and FFLO condensates. Recent improvements~\cite{NicHemLau12a} in spatial resolution of local probes of current may be used to detect and characterize the structure of the superposition the FFLO condensate. Such high-resolution local probes could also be useful in detecting the counterflow in the two surfaces. The spatial modulations would also have optical signatures. For instance, for an FFLO condensate with a sinusoidal modulation (similar to the Larkin-Ovchinikov state in a superconductor), optical scattering experiments should be able to see changes in the reflectivity or refraction index of the surface~\cite{YanKorOre11a} as the wavelength of the incoming light is tuned through the wavelength of the sinusoidal modulations.

Since the vortices are charged, the vortex flow is accompanied by a charge current, which could be detected by transport or noise measurements. Vortices would flow in response to a temperature gradient, and would therefore contribute to the thermopower. %A vortex also has a spin texture which couples to asymmetric Zeeman field and can therefore be pinned by it. 
A vortex-antivortex pair has two zero-energy states and behaves as a two level system. Moreover, a pair carries an integral charge and has no vorticity. Therefore, the interaction between the pairs is Coulomb, and they must be deconfined at $T>0$, contributing to heat transport.
Finally, the midgap spectrum of vortices~\cite{Ser08a} can contribute to the specific heat with a distinct temperature dependence. 

I note that the mean-field study presented here tends to overestimate the critical temperature in a two-dimensional system, which is instead determined by a Kosterlitz-Thouless transition. But it is expected to capture the qualitative structure of the phase diagram. Going beyond the mean-field approximation, determining the structure of the FFLO superposition, the effects of disorder, and the midgap spectrum of FFLO vortices are interesting problems for future studies.

I would also like to note that I recently became aware of Ref.~\onlinecite{EfiLoz11a} that suggests a FFLO state may be realized in graphene double layer structures. However, the gap equation, the extended phase diagram, and the experimental signatures suggested here are not discussed by these authors. Also, graphene is topologically trivial and does not support fractionally charged vortices.~\cite{SerWebFra08a}

%\vspace{1mm}
%\acknowledgements
%\vspace{-3mm}
\acknowledgements
I acknowledge useful discussions with R. Budakian, H. Fertig, L. Li, and M. Parish, and the hospitality of the Aspen Center for Physics where parts of this work were performed. This research was supported by the College of Arts and Sciences at Indiana University, Bloomington.

%\begin{widetext}
%TEXT
%\end{widetext}

\vspace{-3mm}

%-----------------

\end{document}